\pdfoutput=1
\documentclass[10pt,conference,letterpaper]{IEEEtran}

\usepackage[noadjust]{cite}
\usepackage[T1]{fontenc}
\usepackage[utf8]{inputenc}
\usepackage{balance}
\usepackage{url}
\usepackage{subfig}
\usepackage{tabularx}
\usepackage{hhline}
\usepackage{colortbl}  

\newcommand{\docauthor}{Jonathan Will, Lauritz Thamsen, Dominik Scheinert, Jonathan Bader, and Odej Kao}
\newcommand{\docsubject}{Technische Universit\"at Berlin}
\newcommand{\dockeywords}{Scalable Data Analytics, Distributed Dataflows, Runtime Prediction, Resource Allocation, Cluster Management}
\newcommand{\doctitle}{C3O: Collaborative Cluster Configuration Optimization for Distributed Data Processing\\ in Public Clouds}

\PassOptionsToPackage{bookmarks=false}{hyperref}
\usepackage[pdftex]{hyperref}
\hypersetup{
    pdfborder={0 0 0},
    pdfauthor=\docauthor,
    pdftitle=\doctitle,
    pdfsubject=\docsubject,
    pdfkeywords=\dockeywords
}

\usepackage{amsmath,amssymb,amsfonts}
\usepackage{algorithmic}
\usepackage{graphicx}
\usepackage{textcomp}
\usepackage[pdftex,dvipsnames]{xcolor}  
\usepackage{tikz}

\newcommand\copyrighttext{%
  \footnotesize \textcopyright 2021 IEEE. Personal use of this material is permitted.
  Permission from IEEE must be obtained for all other uses, in any current or future
  media, including reprinting/republishing this material for advertising or promotional
  purposes, creating new collective works, for resale or redistribution to servers or
  lists, or reuse of any copyrighted component of this work in other works.
  DOI: \href{https://doi.org/10.1109/IC2E52221.2021.00018}{https://doi.org/10.1109/IC2E52221.2021.00018}}
\newcommand\copyrightnotice{%
\begin{tikzpicture}[remember picture,overlay]
\node[anchor=south,yshift=10pt] at (current page.south) {\fbox{\parbox{\dimexpr\textwidth-\fboxsep-\fboxrule\relax}{\copyrighttext}}};
\end{tikzpicture}%
}

\definecolor{darkgray}{rgb}{0.47, 0.52, 0.6}
\definecolor{dimgray}{rgb}{0.41, 0.41, 0.41}

\newenvironment{gy}{\color{black}}

\newenvironment{bl}{\color{black}}

\newcolumntype{a}{>{\hsize=.25\hsize}X}
\newcolumntype{b}{>{\hsize=.50\hsize}X}
\newcolumntype{d}{>{\hsize=.75\hsize}X}
\newcolumntype{e}{>{\hsize=1.25\hsize}X}

\newcommand*\circled[1]{\tikz[baseline=(char.base)]{
            \node[shape=circle,draw,inner sep=1.5pt] (char) {#1};}}

\def\BibTeX{{\rm B\kern-.05em{\sc i\kern-.025em b}\kern-.08em T\kern-.1667em\lower.7ex\hbox{E}\kern-.125emX}}

\usepackage{xargs}  
\usepackage[disable, colorinlistoftodos, textwidth=8.5mm, textsize=scriptsize]{todonotes}
\newcommandx{\reviewone}[2][1=]{\todo[linecolor=yellow,backgroundcolor=yellow!25,bordercolor=yellow,#1]{#2}}
\newcommandx{\reviewtwo}[2][1=]{\todo[linecolor=blue,backgroundcolor=blue!25,bordercolor=blue,#1]{#2}}
\newcommandx{\reviewthree}[2][1=]{\todo[linecolor=OliveGreen,backgroundcolor=OliveGreen!25,bordercolor=OliveGreen,#1]{#2}}
\newcommandx{\reviewfour}[2][1=]{\todo[linecolor=Plum,backgroundcolor=Plum!25,bordercolor=Plum,#1]{#2}}

\begin{document}

\title{\doctitle}

\author{%
\IEEEauthorblockN{\docauthor}
\IEEEauthorblockA{\docsubject, Germany\\
\{will, lauritz.thamsen, dominik.scheinert, jonathan.bader, odej.kao\}@tu-berlin.de
}}

\maketitle
\copyrightnotice

\begin{abstract}
Distributed dataflow systems enable data-parallel processing of large datasets on clusters.
Public cloud providers offer a large variety and quantity of resources that can be used for such clusters.
Yet, selecting appropriate cloud resources for dataflow jobs –– that neither lead to bottlenecks nor to low resource utilization –– is often challenging, even for expert users such as data engineers.

We present \textit{C3O}, a collaborative system for optimizing data processing cluster configurations in public clouds based on shared historical runtime data.
The shared data is utilized for predicting the runtimes of data processing jobs on different possible cluster configurations, using specialized regression models.
These models take the diverse execution contexts of different users into account and exhibit mean absolute errors below 3\% in our experimental evaluation with 930 unique Spark jobs.

\end{abstract}

\IEEEpeerreviewmaketitle


\section{Introduction}\label{sec:INTRO}
Distributed dataflow systems like Apache Spark~\cite{spark} and Flink~\cite{flink} simplify developing scalable data-parallel programs, reducing especially the need to implement parallelism and fault tolerance.
However, it is often not straightforward to select resources and configure clusters for efficiently executing such programs~\cite{perforator, Lama_AROMA_2012}.
This is the case especially for users who only infrequently run large-scale data processing jobs and without the help of systems operations staff.
For instance, today, many organizations have to analyze large amounts of data every now and then.
Examples are small to medium-sized companies, public sector organizations, and scientists.
Common application areas are bioinformatics, geosciences, or physics~\cite{bux2013parallelization,pegasus_evo}.

The sporadic nature of many data processing use cases makes using public clouds substantially cheaper when compared directly to investing in private cloud/cluster setups.
In cloud environments, especially public clouds, there are several virtual machine types with different hardware configurations available.
Therefore, users can select the most suitable machine type according to their needs.
In addition, they can choose the horizontal scale-out, avoiding potential bottlenecks and significant over-provisioning for their workload.

Most users will also have expectations for the runtime of their jobs.
However, estimating the performance of a distributed data-parallel job is difficult, and users typically overprovision resources to meet their performance targets.
Yet this happens often at the cost of overheads, which increase with larger scale-outs.

Many existing approaches iteratively search for suitable cluster configurations~\cite{cherrypick, hsu2018micky, hsu2018arrow, hsu2018scout, tuneful2020}.
Several other approaches~\cite{ernest, aria, shah2019quick, deepRM2016, perforator} build runtime models, which are then used to evaluate possible configurations, including our previous work~\cite{bell, ellis, verbitskiy2018cobell, koch2017smipe, scheinert2021bellamy}.
Here, training data for the runtime models is typically generated with dedicated profiling runs on reduced samples of the dataset, or historical runtime data is used.
That is, these approaches involve significant overhead for testing configurations, or else assume periodic executions.
The problem of testing over\-head is aggravated in public cloud services like Amazon EMR that have cluster provisioning delays of seven or more minutes\footnote{\href{https://amzn.to/3rRbabd}{https://amzn.to/3rRbabd}, accessed April 2, 2021}.

In this paper, we present \emph{C3O}, a resource allocation system with runtime prediction models for collaborative cluster configuration optimization based on shared runtime data. It is an idea derived from analyzing 930 Spark jobs~\cite{will2020towards}.
Since many different users and organizations use the same public cloud resources, they could also collaborate in performance modeling and selecting good cluster configurations.
We expect especially researchers to be willing to share not just jobs, but also runtime metrics on the execution of jobs, in principle already providing a basis for performance modeling.

\vspace{3mm}
\emph{Contributions}. The contributions of this paper are:
\vspace{-1mm}
\begin{itemize}
    \item A system for collaboratively sharing runtime data to learn optimized cluster configurations for distributed data processing in public clouds\footnote{\href{https://github.com/dos-group/c3o}{https://github.com/dos-group/c3o}}\\[-1.7ex]
    \item A runtime predictor for distributed dataflow jobs, which can utilize shared training data that were generated by different users in diverse contexts\footnote{\href{https://github.com/dos-group/c3o/RuntimePrediction}{https://github.com/dos-group/c3o/RuntimePrediction}}\\[-1.7ex]
    \item An evaluation of our runtime predictor, which also quantifies the benefit of collaboratively shared runtime data, compared to traditional single-user approaches\footnote{\href{https://github.com/dos-group/c3o/evaluation}{https://github.com/dos-group/c3o/evaluation}}\\
\end{itemize}

\pagebreak

\emph{Outline}. The remainder of the paper is structured as follows.
Section~\ref{sec:BACKGROUND} explains the background of this work.
Section~\ref{sec:SYSTEM_IDEA} presents the overall system idea for C3O, a system for collaborative sharing of runtime data.
Section~\ref{sec:CLUSTER_CONFIGURATOR} explains in further detail our cluster configuration strategies.
Section~\ref{sec:RUNTIME_PREDICTION} dissects the cluster configurator's key component, its runtime predictor.
Section~\ref{sec:EVALUATION} evaluates the runtime predictor and compares collaborative approaches to single-user approaches.
Section~\ref{sec:RELATED_WORK} discusses related work.
Section~\ref{sec:CONCLUSION} summarizes and concludes this paper.

\section{Background}\label{sec:BACKGROUND}
In this section, systems for analyzing large datasets are reviewed, including distributed file systems, distributed dataflow systems and relevant cloud services.

\subsection{Distributed File Systems}

Distributed file systems are a crucial component of big data analytics.
Large files are split up into equally sized blocks of e.g. 64MB, that get distributed among multiple nodes, typically consisting of commodity hardware.
Each block is stored redundantly among several machines, so-called \emph{Data Nodes} to avoid data loss in the case of hardware failures.

Splitting up the data into blocks allows for saving files of practically unlimited size.
However, access over the network is typically of higher latency than access to traditional, local file systems.
Nonetheless, storing the data redundantly on different machines means that data can be accessed in parallel, resulting in a higher possible aggregate bandwidth compared to non-redundant storage solutions.

In order to access a file, the client has to consult the \emph{Meta Data Node}, also called \emph{Name Node} or \emph{Master Node}.
The Meta Data Node then returns a list of all the locations of blocks belonging to the requested file, which the client can then piece back together to obtain the original file.
For writing new files to the system, the client needs to request a list of designated locations to write the blocks to, which is done in parallel.

\begin{figure}[htb]
    \centering
    \includegraphics[width=1\columnwidth]{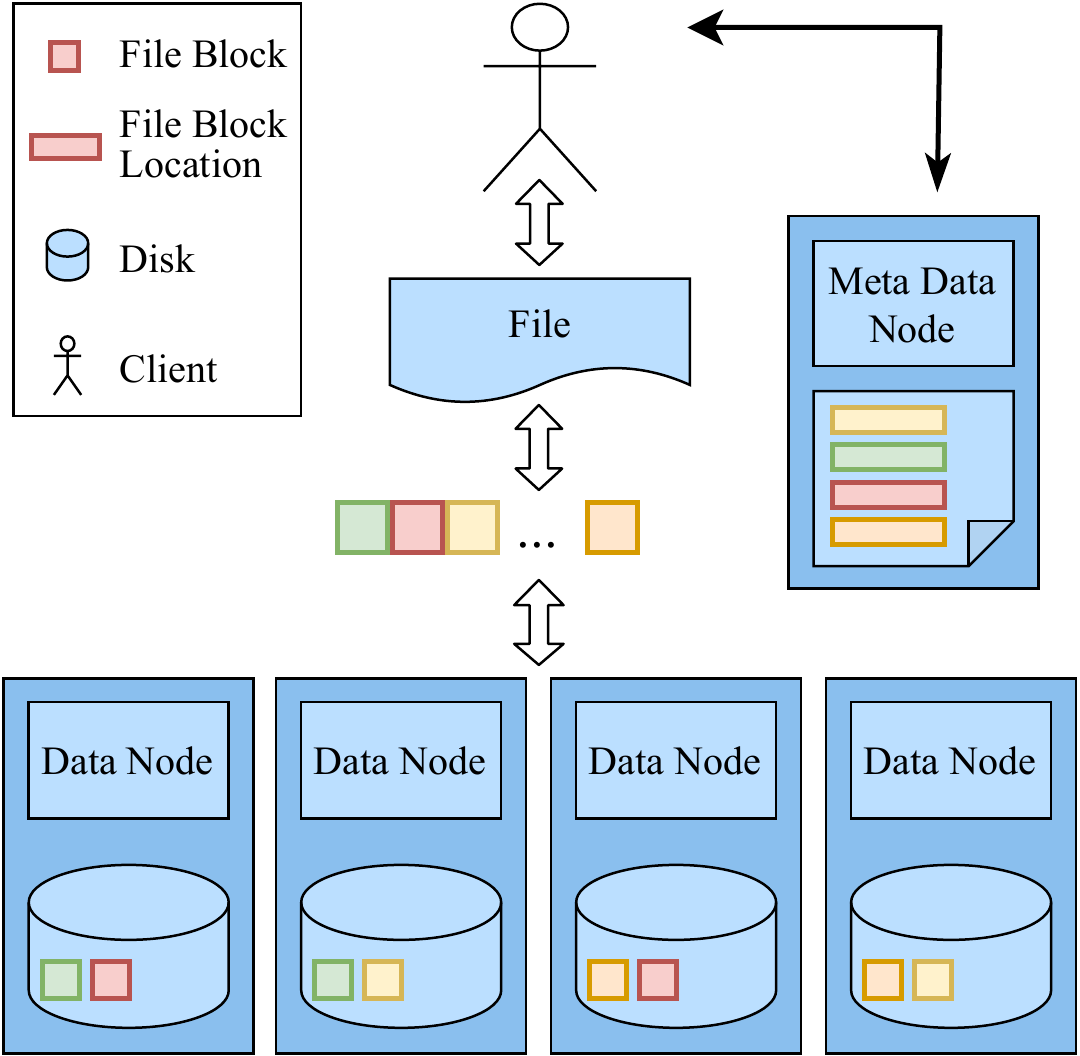}
    \caption{General architecture of distributed file systems}\label{fig:dfs}
\end{figure}

An overview of the components and processes within distributed file systems is depicted in Fig.~\ref{fig:dfs}.

Prominent examples of distributed file systems are the Google File System~\cite{gfs} and the Hadoop File System~\cite{hdfs}, which is an open-source alternative to the former.


\subsection{Distributed Dataflow Systems}

Distributed dataflows are graphs of connected data-parallel operators, which execute user-defined functions on a set of shared-nothing commodity cluster nodes.
Through high-level programming abstractions, users can easily create data-parallel programs without having to explicitly handle the parallelization.
It is the system that translates the sequential program of the user into a directed graph of data-parallel operators and finally into an optimized execution plan.
Error handling is also taken care of by such systems.
Failed operations are repeated and defective nodes are replaced by intact ones.

\begin{figure}[hbt]
    \centering
    \includegraphics[width=1.0\columnwidth]{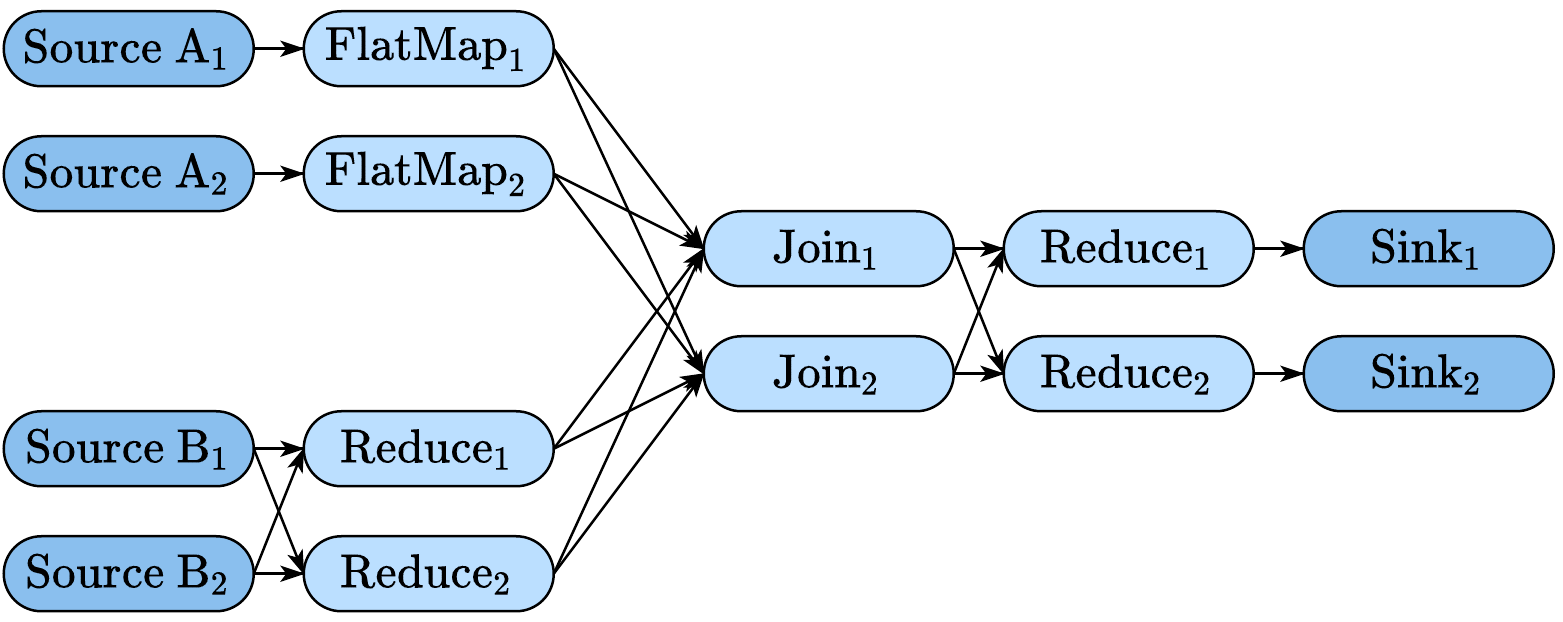}
    \vspace{5mm}
    \caption{Example of a parallelized dataflow graph}\label{fig:dataflows}
\end{figure}

Fig.~\ref{fig:dataflows} shows an example of a distributed dataflow program represented by a graph of data-parallel operators.
Here, two parallel task instances operate on two partitions of the data.
Operators that just process the data points individually, like \emph{FlatMap}, can be executed on partitions of the dataset in parallel, thus leading to a speed-up.
On the other hand, operators that require a full view of the data, e.g. \emph{Reduce} have synchronization barriers.

Specific examples of distributed dataflow systems are Hadoop MapReduce~\cite{mapreduce}, Apache Flink~\cite{flink}, and Apache Spark~\cite{spark}, the latter two of which can be seen as an evolutionary step up from the former.

\subsection{Co-located Analytics Clusters in Clouds}

When built on top of a distributed file system, distributed dataflow systems form co-located analytics clusters, i.e.\ the nodes that hold the data are used for computation during the execution of a distributed dataflow job.
The underlying distributed file system can then be used to initially read the dataset in parallel, save intermediate results and recovery information, and write back the fully processed data.
Here, distributed dataflow systems can gain performance increases by considering data locality when deciding what data to process on which node.

Cloud computing is the on-demand availability of computer system resources, without direct active management by the user.
The services are accessible completely over the network.
Users can benefit greatly from this if they have sporadic, ever-changing processing needs or if they lack the capital or know-how to invest in their own dedicated cluster.
Popular public cloud services with such offerings include Amazon Web Services (AWS)\footnote{\href{https://aws.amazon.com/}{https://aws.amazon.com}, accessed March 31, 2021}, Google Cloud\footnote{\href{https://cloud.google.com/}{https://cloud.google.com}, accessed March 31, 2021}, and IBM cloud\footnote{\href{https://www.ibm.com/cloud}{https://www.ibm.com/cloud}, accessed March 31, 2021}.
Cloud services provide two main functionalities: Storage and computation.
Both are essential for large-scale data analytics.

While co-location of storage and compute is typical in dedicated clusters that often consist of commodity hardware, this is not practical in public clouds due to the on-demand nature of access and payment per minute.
Users that keep a large dataset stored in a public cloud, typically do so via long-term storage services instead of reserving full nodes and in turn paying for the unused compute power.
An example of such a service is \textit{Amazon S3}\footnote{\href{https://amzn.to/31SkGR2}{https://amzn.to/31SkGR2}, accessed April 2, 2021} on AWS, a storage solution in which objects are organized and stored redundantly within so-called ``buckets''.

In practice, the co-location of the data and computation on the same cluster of nodes is therefore only realized once a job is to be executed.
After the execution, any processed data can be written back to long-term storage.
The cluster can then be torn down immediately, incurring no further costs to the user.
Major public cloud services offer managed cluster platforms that provide users with an HDFS cluster and the ability to submit jobs of various distributed dataflow systems to it.

The user can choose to provision cluster resources from a variety of machine types, specializing in different domains, like memory size and I/O speeds.
Also, the amount of master and worker nodes can be chosen at the user's discretion.
Any necessary software, such as the data processing system and the HDFS comes pre-installed on all nodes of the cluster.
After a few minutes of startup time, the cluster is in a fresh state and ready to process any submitted jobs.

\begin{figure}[htb]
    \centering
    \includegraphics[width=1\columnwidth]{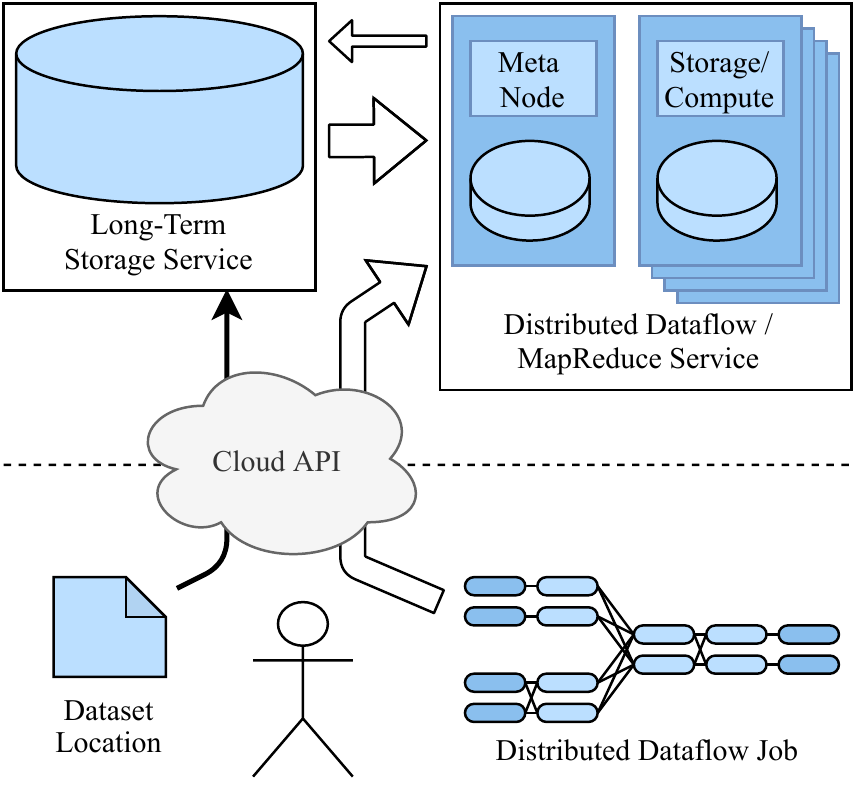}
    \caption{General functioning of co-located analytics clusters in clouds}\label{fig:clacc}
\end{figure}

Fig.~\ref{fig:clacc} gives an overview of how co-located analytics clusters in clouds function.

An example of such a managed cluster service on a public cloud is \textit{Amazon EMR}\footnote{\href{https://amzn.to/3rWq4gi}{https://amzn.to/3rWq4gi}, accessed April 2, 2021} on AWS\@.
Its integration with the long term storage Amazon S3 is accomplished through the so-called \textit{EMRFS}. It enables the job to read and write directly to and from Amazon S3 as if it were part of the underlying HDFS.

\pagebreak

\section{System Idea}\label{sec:SYSTEM_IDEA}
This section presents our approach to the problem of finding the best cluster configuration for a distributed dataflow job.
We first present the overall idea of a collaborative runtime prediction system, building upon our previous work~\cite{will2020towards, bell}, and then explain a possible system architecture for an implementation of the approach.

\subsection{Main Idea}

Especially with open source software, users share implementations of common jobs and algorithms instead of implementing these themselves.
Many of the most common distributed dataflow jobs are therefore being run every day by different individuals or organizations worldwide.
Consequently, the runtime data resulting from these executions could be shared for the benefit of all, allowing for accurate runtime predictions from the first execution of a job in an organization.
That is, the main idea of a collaborative optimization of cluster configurations is to share historical runtime data alongside the code for the jobs and prediction models, which allow users to benefit from global knowledge in both efficient algorithms and cluster configuration simultaneously.
Just like the users can contribute code to the repository in which they found the program they are using, they can also contribute their runtime data.

The code contributors to such repositories, henceforth called \emph{maintainers}, can use their domain knowledge to fine-tune the default models that come with the system to suit the job at hand or add entirely new, specialized models to it.
All users then benefit from this.

In\reviewthree{[3.2]} the end, users of this system can easily find implementations of at least somewhat common data analytics jobs, which also include runtime data and runtime models.
After selecting an implementation, they can provide their own dataset and parameters and execute the job on an automatically selected resource-efficient and performance target fulfilling cloud configuration.
All of this can happen in one seemless process.

\subsection{Architecture \& Operation}

\begin{figure}[htbp]
    \centerline{\includegraphics[width=\columnwidth]{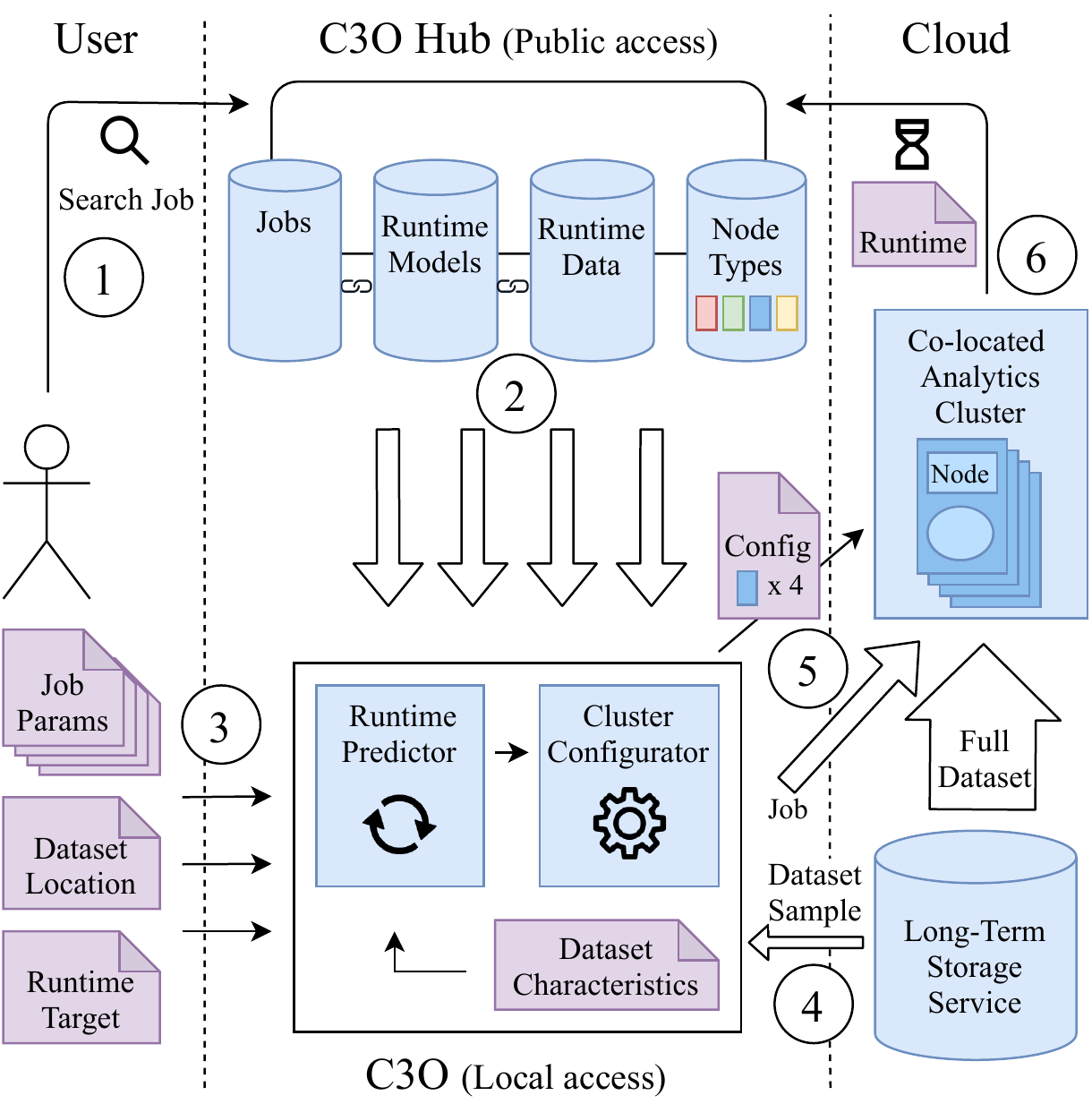}}
\caption{Configuring and using a public cloud cluster for distributed data processing via C3O}\label{fig:figure_one}
\end{figure}

The overall system contains the runtime prediction and cluster configuration system C3O.
Besides, there is a website called \emph{C3O Hub}.
Fig.~\ref{fig:figure_one} illustrates the system architecture and depicts the envisioned workflow for the users.

\circled{1} Users start by looking for a job that implements the data analytics algorithm that they want to execute on their dataset.
The repositories containing the code and the associated runtime data can be found by users on C3O Hub, which lists them along with meta-information, especially the underlying algorithm.

\circled{2} The cluster configuration system, C3O, downloads the desired job, associated runtime data, and potentially any associated job-specific runtime models.
Hereby the system also retrieves a list of available virtual machine types on public clouds and their current prices.

\circled{3} Users can now provide job inputs in the form of job parameters, a dataset location, and optionally a runtime target.

\circled{4} The last information that the \emph{runtime predictor} requires is the key metadata about the dataset to be processed.
This information is extracted from a small sample of the dataset.
In\reviewtwo{[2.3]} general, both the size of the dataset and the data format are important.
The exact configuration of this depends on the given job and is the responsibility of the maintainers.

\circled{5} According to the runtime predictions for each possible configuration and the runtime target, the \emph{cluster configurator} chooses the most suitable cluster configuration.
This cluster of a specified node type and scale-out is\reviewone{[1.10]} then reserved in a public cloud and used to run the job.

\circled{6} Finally, after the job has been executed, the newly generated runtime data is captured and saved.

Altogether, the components form a system that streamlines the process of executing a distributed dataflow job on a user's data, as well as configure and create a cluster that fulfills the user's performance constraints.

\subsection{Collaboration}

Users collaborate by sharing their historic runtime data.
Besides the default models of C3O, maintainers can include custom runtime models that capture the runtime behavior of the particular job at hand.

\paragraph{Sharing Runtime Data}

We implement the technical aspect of sharing historical runtime data alongside the code of a distributed dataflow job by putting both into the same code repository.
A downside to this rather simple approach is the code commit history being diluted by data commits.

Another way to allow collaboration on runtime data from many users would be to use a dedicated dataset version control system like DataHub~\cite{datahub} and reference it from the code repository.
An alternative is DVC\footnote{\href{https://dvc.org/}{https://dvc.org}, accessed April 09, 2021} which addresses code versioning and dataset versioning simultaneously.
Such systems provide functions like \emph{fork} and \emph{merge}, which are known from code version control systems.


However, the most crucial part of sharing runtime data is to accurately capture the full context of the execution.
The actual features to be collected and shared are determined by the maintainers.
Natural candidates are the job parameters and key dataset characteristics.
The latter can often be determined by looking at a small sample of the dataset prior to execution, like the size and format of the data.

In other cases, a sample can only provide an estimate of a key dataset characteristic.
Then it is necessary to extract the exact value after the execution.
An example here for a Grep job is the number of occurrences of the keyword in the dataset.

\paragraph{Validating Shared Data}
\reviewfour{[4.2]}\reviewthree{[3.3]}A necessary step between receiving runtime data from users and publishing it in a shared repository is to validate the data.
This data might be either inadvertendly corrupted or fabricated data might be submitted with malicious intent.
In either case, the prediction accuracy of the runtime models would be compromised when using this as training data.

A possible solution to the issue is to retrain the prediction models while incorporating the new training data and then evaluating the runtime predictor accuracy on a test dataset consisting of previously existing datapoints.
Should the evaluation exhibit a significant increase in prediction errors, then the new runtime data contribution will be rejected.

\paragraph{Sharing Custom Runtime Models}

Maintainers can add custom, job-specific runtime models for the runtime predictor in an effort to increase its runtime prediction accuracy.
These models can be included in the same code repository as the job.
To integrate all the models into the overall runtime predictor, it is important that they all share a common API\@.

\subsection{Selecting a Cluster Configuration}

To find a suitable cluster configuration, we train regression models that learn expected runtimes based on previous executions of the same job.
The  runtime estimates for different scale-outs are used to choose a cluster size that is in accordance with the user's expectation of runtime and cost of execution.

In a traditional single-user scenario, with only locally generated runtime data, runtime models can succeed with only learning few runtime-influencing factors.
Most importantly, the different scale-outs and the size of the dataset have to be considered.
With enough training data, the regression models can then interpolate or extrapolate to different dataset sizes and different scale-outs.

In a collaborative setting, runtime metrics produced globally by different users can be expected to vary in more than just the previously mentioned runtime influencing factors.
While a single user might always choose the same value for a runtime-influencing algorithm parameter, different users will choose different values according to their individual context.
An example here is \emph{k} in the K-Means algorithm.
Further, important dataset characteristics that have an influence on the runtime are expected to differ for different users.
For instance, two datasets for PageRank can have the same size in megabytes and the same amount of links, while containing vastly different amounts of unique pages.
This would result in significant disparities in the problem size and thus the resulting runtimes.

An increasingly high feature space dimensionality renders available training data sparse.
On the other hand, through the global sharing of the runtime data, users will also have more training data at their disposal.
In this setting, our predictor takes the approach of choosing the most promising model out of several based on cross-validation with the available training data.

The two key components of our system, the cluster configurator and the runtime predictor, shall be examined more closely in the following two sections.

\section{Cluster Configurator}\label{sec:CLUSTER_CONFIGURATOR}
In this section, we present the C3O cluster configurator\footnote{\href{https://github.com/dos-group/c3o/ClusterConfiguration}{https://github.com/dos-group/c3o/ClusterConfiguration}}.
Configuring a cluster in our context entails selecting a machine type out of the numerous virtual machines being offered by the chosen public cloud provider and specifying a scale-out, i.e.\ the number of nodes.

Since the related work~\cite{hsu2018arrow} and our own experimental data~\cite{will2020towards} indicated that the optimal machine type is usually job-dependent and scale-out-independent, the choices for machine type and scale-out will be made successively.

\subsection{Selecting a Suitable Machine Type}

One objective of cluster configuration is to find the most resource-efficient machine type for the problem at hand.
Different algorithms have different resource needs regarding CPU, memory, disk I/O and network communication.
An efficient machine fulfills all those needs, avoiding hardware bottlenecks, while also not overfulfilling any of them.

We can calculate the overall cost of a job on different machine types by multiplying the machine type's operating cost, the execution time, and the chosen scale-out.

In the context of collaborative cluster configuration, the maintainer of a C3O repository is expected to find a suitable machine type based on the job at hand by doing test runs on different machine types.
Once a good machine type is found, users can use this and then adjust their scale-out based on their individual problem, meaning their specific job inputs and according to their needs.
As a fallback in the case where the maintainers have not yet selected a machine type, the system preferably chooses a general-purpose machine for which there is runtime data available.

\subsection{Selecting a Suitable Scale-Out}

While the machine type can already be chosen based solely on the job type, the scale-out needs to be selected carefully, ensuring fulfillment of the user's runtime and cost preferences.

Having accurate runtime predictions is essential for selecting a suitable scale-out.
We can assume that every job has some form of deadline or at least an expectation towards its runtime.
This can range anywhere from minutes to days.
Thus, assuming there is a maximum allowed runtime $t_{max}$ given by a deadline, C3O chooses the scale-out $\hat{s}$ to be the smallest available scale-out $s$ with estimated runtime $t_s$ and runtime prediction error $\epsilon$ that is expected to fulfill that deadline with the specified confidence~$c$. Formally:


\begin{equation*}
\hat{s} \; = \; \text{min}\big\{ \, s \in S \; | \; P(t_s + \epsilon \le t_{max}) \ge c \, \big\}
\end{equation*}

The confidence value can be configured by the user, with the default of C3O being $0.95$.
We can solve the above equation by calculating a value $\epsilon_c$ for which the actual error $\epsilon$ is not higher, with a chance of $c$ given by $c = P(\epsilon \le \epsilon_c)$, which then leads to:

\begin{equation*}
\hat{s} \; = \; \text{min}\big\{ \, s \in S \; | \;  t_s + \epsilon_c \le t_{max}  \, \big\}
\end{equation*}

We\reviewtwo{[2.6]}\reviewone{[1.2]}\reviewone{[1.6]} examined the error distribution for all combinations of models and jobs and found that the majority exhibited a Gaussian error distribution\footnote{\href{https://github.com/dos-group/c3o/evaluation}{https://github.com/dos-group/c3o/evaluation}}.
Thus, assuming a Gaussian distributed runtime prediction error $\epsilon \sim \mathcal{N(\mu, \sigma^{\text{2}})}$, we can calculate $\epsilon_c$ by utilizing the mean and standard deviation of the errors that resulted from the cross-validation of the model that predicted the runtimes.

The ratio of points that lie within the interval $\mu \pm x\sigma$ is given by
$ {\displaystyle \operatorname {erf} ({x / \sqrt {2}})} $, with erf being the \textit{Gauss error function}.

We are only interested in the ratio of prediction errors that are \textit{not too high} to fit into this range,
thereby excluding the errors that came from significant runtime underestimations.
Consequently, the ratio $c$ should be $c = P(\epsilon \le \mu + x\sigma)$.
We are thus looking for an $x$, so that $1-c$ is the ratio of points with values that are too large for the interval, and therefore $2 \cdot(1 - c)$ is the total ratio of points outside of the interval and $1 - 2 \cdot(1 - c) = 2c -1$ is the ratio of points inside the interval:

\begin{equation*}
\displaystyle \operatorname{erf} (x / \sqrt{2}) \quad \overset{!}{=} \quad 2c -1
\end{equation*}

Using the \textit{inverse Gauss error function}, we solve for $x$ with

\begin{equation*}
x = \operatorname{erf}^{-1}(2c-1) \cdot \sqrt{2},
\end{equation*}

such that our complete equation eventually summarizes to

\begin{equation*}
\hat{s} \; = \; \text{min}\big\{ \, s \in S \; | \;  t_s +
\left( \mu + (\operatorname{erf}^{-1}(2c-1)\cdot\sqrt{2})\sigma \right)
\le t_{max}  \, \big\}.
\end{equation*}

For instance, if the confidence value is set to $0.95$, then valid options are scale-outs with predicted runtimes $t_s$ so that:

\begin{equation*}
t_s + \mu + 1.64485 \cdot \sigma \le t_{max} \quad \text{(Rounded value)}.
\end{equation*}

True and hard deadlines are expected to occur mostly with jobs that are recurring organization-internally.
One-off jobs can still benefit from a careful choice of the scale-out, e.g.\ due to the aforementioned possible hardware bottlenecks at certain scale-outs, leading to situations where the lowest possible scale-out is not the cheapest option for the job execution.
An insufficient scale-out can lead to a dataset not fitting into the total combined memory of the cluster.
This is a problem especially for iterative algorithms where the dataset has to be read from and written back to disk for each iteration, instead of remaining in memory.
The result is massive runtime increases over sometimes only slightly higher scale-outs.

The system will not recommend configurations where a hardware bottleneck can be expected, like scale-outs that are too low, unless there is no valid other option without such an expected bottleneck.

In cases where runtime and cost are of equal concern, the users are presented pairs of estimated runtimes and resulting prices, each pair corresponding to an available scale-out.
This can also be visualized to the users through plots, which enables them to quickly make an informed decision.

\section{Runtime Prediction}\label{sec:RUNTIME_PREDICTION}
To find a suitable cluster configuration, we train regression models that learn expected runtimes based on previous executions of the same job.
The  runtime estimates for different scale-outs are used to choose a cluster size that is in accordance with the user's expectation of speed and cost of execution.
In the following, we present these runtime models and our strategy of employing them to make runtime predictions.

\subsection{General Models}

Depending on the job in question, the feature space can grow quite large, while the amount of available training data may differ considerably.
However, our system requires models that can facilitate runtime prediction for all possible distributed dataflow jobs with satisfactory accuracy.

To\reviewthree{[3.4]} choose the general models that come with the system, we look at the average performance of the models accross all jobs.
This can be weighted by the commonness of the jobs.
Those models with the lowest overall errors provide a solid basis for each newly created C3O repository.

One such model is \emph{gradient boosting} (GBM), which we found to be a formidable choice for most of our examined jobs and scenarios.
It is an ensemble method where the predictions of many so-called ``weak learners'' are combined into one final prediction. The individual weak learners are discovered in a sequential manner, with each one trying to correct the errors of its predecessor.
This technique can therefore succeed almost regardless of feature-dimensionality and interdependence of features.

\subsection{Custom Models}

Considering the characteristics of the job at hand, the maintainers of a C3O repository can create their own custom runtime models that can more closely capture the behavior of this job, further increasing prediction accuracy.

For the jobs in our evaluation dataset, most of the main runtime-influencing factors appear to be pairwise independent.
Based on these observations, our strategy is to learn the influence of pairwise independent features and then finally recombine those models.
This results in several models of low-dimensional feature spaces.
Owing to the curse of dimensionality, these together require less dense training data than single models that consider all features simultaneously.
Because of this optimistic assumption that runtime-influencing factors are pairwise independent, we call it the \emph{optimistic approach}.

Models based on our optimistic approach consist of a \emph{scale-out to speedup model} (SSM) and an \emph{inputs behavior model} (IBM).
The SSM shall be trained on training data points that share the same values for every feature except for the scale-out.
After the SSM has been trained, it is used to project all data points onto a scale-out of 1.
With these transformed training data points, the IBM is trained.
Predictions are then made by feeding a data point into both models and multiplying the results.

Specifically, we implemented one \textit{basic optimistic model} (BOM), which simply uses linear regression for its IBM and a model based on a third-degree polynomial for its SSM\@.
Further, we created a model we call \textit{optimistic gradient boosting} (OGB), which uses gradient boosting for both the IBM and the SSM\@.

\subsection{Dynamic Model Selection}

Which of the models performs best depends on the par\-ticular situation.
For a given job, the specific implementations of the available models influences their prediction accuracy.
Also, the quantity and quality of available training data points are important factors to be considered.
While some models can initially make comparatively good predictions and not benefit significantly as more training data becomes available, for others, the opposite is the case.

Realistically, training data characteristics change as time progresses and more runtime data is added to the training dataset.
Hence, we switch dynamically between prediction models depending on the expected accuracy.
The models are retrained on the arrival of new runtime data.
Based on cross-validation, the most accurate model averaged over the test datasets is chosen to predict new data points.

\section{Evaluation}\label{sec:EVALUATION}
To test the quality of our cluster configuration system, we implemented the runtime predictor on which its decisions are based.
We then evaluated the predictor and its constituent models on a dataset comprising 930 unique runtime experiments across 5 different Spark jobs, all of which were executed via Amazon EMR on AWS\@.
Specifically, we examined prediction performance in collaborative scenarios and traditional single-user scenarios by managing access to categories of training data points.
Next, we looked at how the prediction accuracies develop with increasing amounts of available training data.
To put the results into perspective, our evaluations include the perhaps most prominent runtime model based cluster configuration approach, Ernest~\cite{ernest}, as a baseline.

\subsection{Prototype Implementation}

For the implementations of both the cluster configurator and the runtime predictor, we chose Python (version 3.9) for its code readability and its wealth of available libraries.
One such library in particular is \textit{Scikit-Learn} (v. 0.24.1)~\cite{scikit-learn}, which we benefited from when building the C3O runtime predictor and its models.
Major supporting libraries we used were \textit{Numpy} (v. 1.20.2), \textit{Pandas} (v. 1.2.4), and \emph{scipy} (v. 1.6.2).
The cluster configurator is implemented as a command-line tool.

We organize our runtime data in a TSV format, containing first the machine type and the instance count, and job-specific context-describing features at the end.

\subsection{Datasets}

In our evaluation, we used the runtime data\footnote{\href{https://github.com/dos-group/c3o-experiments}{https://github.com/dos-group/c3o-experiments}} we published previously~\cite{will2020towards}.
It contains job executions of five different algorithms that were tested under various cluster configurations in Amazon EMR 6.0.0, which uses Hadoop 3.2.1 and Spark 2.4.4.

The JAR files containing those algorithms were compiled with Scala version 2.12.8.

In\reviewone{[1.9]} total, we executed 930 unique runtime experiments, which were each executed five times.
An overview of them can be seen in Table~\ref{table:test_jobs}.

The column ``\#Features'' describes the number of runtime-influencing features in the training data.
Three of these are shared among all jobs:
\begin{itemize}
    \item The machine type
    \item The scale-out
    \item The dataset size / problem size
\end{itemize}
\vspace{1mm}
Additionally, jobs can have features that capture their specific context.
These are related to the algorithm parameters of the job or to key dataset characteristics.
An example of this is the number of \textit{unique} pages in a dataset for a PageRank job.
The Sort dataset has no additional context-describing features and thus there can be no distinction between global and local training data.

Each of the 930\reviewone{[1.9]} runtime experiments was conducted five times, and only the median runtimes are considered here in order to control for possible outliers that might occur through e.g.\ partial hardware failures in the cluster during execution.
The algorithms of the jobs contain only standard implementations that come with the official libraries of Spark.

\begin{table}[h!]
    \centering
    \caption{Overview of Runtime Data for Model Evaluation}
    \begin{tabularx}{\columnwidth}{|d|a|X|X|X|c|}
    \hline
             & Jobs & Datasets                            & Input Sizes & Parameters                                        & \#Features \\ \hhline{|=|=|=|=|=|=|}
    Sort     & 126  & Lines of random chars               & 10-20~GB    & ---                                               & 3+0 \\ \hline
    Grep     & 162  & Lines of random chars and keywords  & 10-20~GB    & Keyword\newline ``Computer''                      & 3+1 \\ \hline
    SGD      & 180  & Labeled Points                      & 10-30~GB    & Max.\ iterations 1-100                            & 3+2 \\ \hline
    K-Means  & 180  & Points                              & 10-20~GB    & 3-9 clusters,\newline convergence criterion 0.001 & 3+2 \\ \hline
    PageRank & 282  & Graph                               & 130-440~MB  & convergence criterion\newline 0.01-0.0001         & 3+2 \\ \hline
    \end{tabularx}
    \label{table:test_jobs}
\end{table}

\subsection{Experiments}

Evaluating the quality of our solution, we examined prediction performance in collaborative scenarios and traditional single-user scenarios, as well as how the amount of available training data influences the model accuracies.

Since\reviewtwo{[2.2]}\reviewone{[1.8]} the C3O runtime predictor chooses its runtime model based on cross-validation, the number of considered train-test splits leads to a trade-off between the selection accuracy and the overhead of repeatedly training the models.
In the implementation used for the evaluation, \emph{leave-one-out} cross-validation is used, which lead to a maximum, yet manageable runtime of 10-30 seconds for model selection.
With increasing training datasets, the model selection phase needs to be capped, either by setting a time budget or limiting the number of train-test splits used in the cross-validation.

In all of these experiments, the models only learned from training data that was generated on the target machine type. 
This is also in line with the assumption that the suitability of a machine type is stable based on the job and is in line with the sequential nature of machine type selection and scale-out selection as discussed in Section~\ref{sec:CLUSTER_CONFIGURATOR}.

\paragraph{Training Data Origin}

\begin{table*}[h!]
    \caption{Runtime Prediction Accuracy of Different Models and the C3O Predictor When Considering Local-Only or Globally Created Training Data: Mean Absolute Percentage Error}
    \begin{tabular}{l@{\hspace{2mm}}r!{\vrule}}
        \textbf{Sort}\\[0.8ex]
               & local/global\\
        \hline\hline
        Ernest & \bl{5.82\%} \\
        GBM    & \bl{4.43\%} \\
        BOM    & \bl{6.39\%} \\
        OGB    &  \textbf{\bl{2.61\%}} \\[-0.3ex]
        \arrayrulecolor{lightgray}\hline
        C3O    &  \textbf{\bl{2.61\%}} \\
    \end{tabular}
    \hspace{0mm}
    \begin{tabular}{l@{\hspace{3mm}}rr!{\vrule}}
        \textbf{Grep}\\[0.8ex]
               & local & global \\
        \hline\hline
        Ernest & \bl{7.53\%} & \gy{39.38\%} \\
        GBM    & \gy{5.54\%} & \textbf{\bl{ 2.74\%}} \\
        BOM    & \bl{6.45\%} & \gy{12.95\%} \\
        OGB    & \textbf{\bl{4.47\%}} & \gy{ 9.35\%} \\[-0.3ex]
        \arrayrulecolor{lightgray}\hline
        C3O    & \gy{5.05}\% & \textbf{\bl{ 2.74\%}} \\
    \end{tabular}
    \hspace{0mm}
    \begin{tabular}{l@{\hspace{1mm}}rr!{\vrule}}
        \textbf{SGDLR}\\[0.8ex]
               & local & global \\
        \hline\hline
        Ernest &         \bl{10.00\%}  &         \gy{21.85\%} \\
        GBM    &         \gy{ 6.89\%}  & \textbf{\bl{ 2.25\%}} \\
        BOM    & \textbf{\bl{ 6.04\%}} &         \gy{12.66\%} \\
        OGB    &         \bl{ 6.54\%}  &         \gy{ 7.79\%} \\[-0.3ex]
        \arrayrulecolor{lightgray}\hline
        C3O    &         \gy{ 6.22}\%  & \textbf{\bl{ 2.25}}\% \\
    \end{tabular}
    \hspace{0mm}
    \begin{tabular}{l@{\hspace{1mm}}rr!{\vrule}}
        \textbf{K-Means}\\[0.8ex]
               & local & global \\
        \hline\hline
        Ernest & \bl{14.04\%} & \gy{15.31\%} \\
        GBM    & \gy{ 8.60\%} & \textbf{\bl{ 2.17\%}} \\
        BOM    & \bl{ 5.51\%} & \gy{ 5.74\%} \\
        OGB    & \gy{ 5.70\%} & \bl{ 5.50\%} \\[-0.3ex]
        \arrayrulecolor{lightgray}\hline
        C3O    &  \textbf{\gy{5.22\%}} & \textbf{\bl{2.17\%}} \\
    \end{tabular}
    \hspace{0mm}
    \begin{tabular}{l@{\hspace{-1mm}}rr!{\vrule}}
        \textbf{Page Rank}\\[0.8ex]
               & local & global \\
        \hline\hline
        Ernest & \bl{10.93\%} & \gy{34.85\%} \\
        GBM    & \gy{ 5.25\%} & \textbf{\bl{ 2.71\%}} \\
        BOM    & \textbf{\bl{ 3.99\%}} & \gy{15.08\%} \\
        OGB    & \gy{ 4.05\%} & \bl{ 3.17\%} \\[-0.3ex]
        \arrayrulecolor{lightgray}\hline
        C3O    & \gy{4.29\%} &  \bl{ 2.77\%} \\
    \end{tabular}
    \label{tab:categories}
\end{table*}

First, we want to evaluate the collaborative approach against the traditional non-collaborative approach.

We emulate the traditional single-user situation by only providing models with \textit{local} training data that can be regarded as all belonging to the same context.
This means that while scale-outs and dataset sizes are still variable, other runtime-influencing dataset characteristics and the algorithm parameters to the job are the same for each data point in the training dataset.
For instance, runtime data from a training dataset for Grep will have the same ratio of lines containing the keyword and a training dataset for K-Means will only contain data points that all have the same \textit{k}.
That way, multiple valid local training datasets exist and the train-test splits for the evaluation experiments are chosen uniformly from those.

On the other hand, \textit{global} training data is supposed to represent runtime data that also stems from other collaborating users, in potentially different contexts.
Therefore it varies in all features, but in turn, it makes those training datasets larger.

One special case here is Sort, where the local and global training datasets were the same, since its only features are the dataset size and the scale-out.

For every job, each of the models, as well as the C3O predictor as a whole were evaluated on 300 train-test splits.
We recorded the averages of the mean absolute percentage errors.

Table~\ref{tab:categories} shows the results of this evaluation.
The BOM and OGB tend to perform better when they are trained only with data points that stem from the same context.
Due to the way the optimistic models are built, they can perform well if they only need to learn the impact of scale-out and dataset size, while being less accommodating to additional features.
The GBM on the other hand appears to benefit from the additional training data enough to see an increase in accuracy in the collaborative scenario in spite of the increased number of features to be learned.

Since Ernest was not built to consider any features other than the dataset size and the scale-out at all, it is only suitable for single-user scenarios with locally generated training data from the same context.
This is evidenced by its considerably reduced performance in the collaborative scenario, when trained with global data from different contexts.
Ernest can however still serve to provide a comparison in our local-only scenarios and for Sort, which is a job that does not require consideration of any wider context.
Here, our novel runtime models, and in turn the C3O runtime predictor also exhibit substantially lower errors than Ernest, with mean absolute percentage errors of 2.61\%-6.22\% against Ernest's 5.82\%-14.04\%.

For the majority of cases, the C3O predictor is at least as accurate as its most accurate constituent model.
In the other occurrences, it is roughly within half a percent of the most accurate individual model.

\paragraph{Training Data Availability}

\begin{figure}
\includegraphics[width=\columnwidth, keepaspectratio]{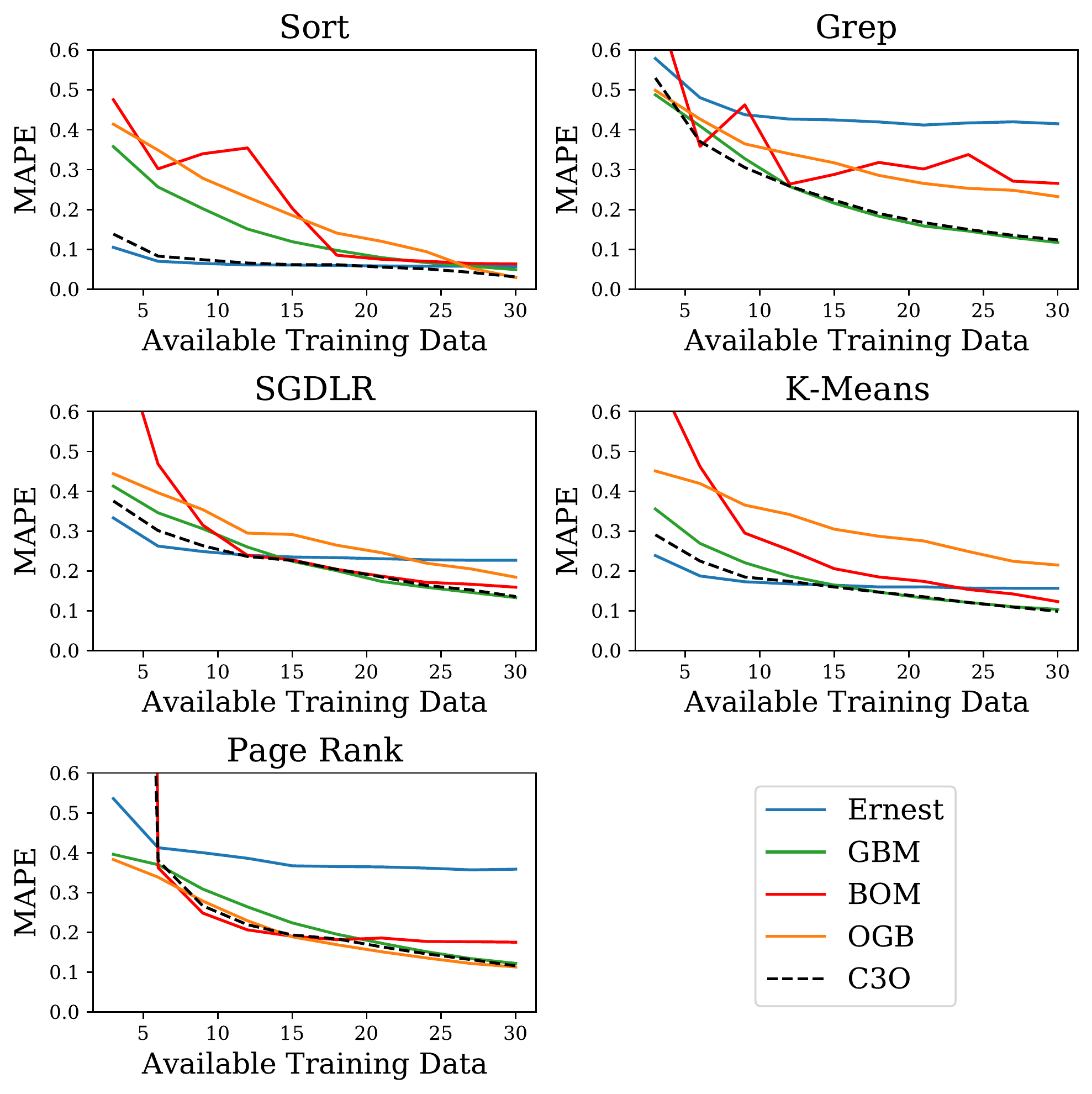}
    \caption{Development of prediction accuracies of different models and the C3O predictor at varying training data availabilities}
    \label{fig:availability}
\end{figure}

The second part of our evaluation experiments examined the development of prediction accuracy for each of the individual models and the C3O predictor as more training data becomes available.

Again, for each of the five jobs, each of the models, as well as the C3O predictor as a whole were evaluated on 300 train-test splits and we recorded the averages of the mean percentage errors.
However, this time train-test splits consisted of 3, 6, \ldots 30 training data points and the remaining data points formed the test set.
The training data was taken from the global training dataset in order to test the models under the conditions of a collaborative setting, with high-dimensional feature spaces but low amounts of available training data.

Fig.~\ref{fig:availability} presents the results of this experiment.
We see that different models can have vastly different convergence behaviors.
Models that perform best at very low training data availability are often not the most accurate when given a larger training dataset.

It is noticeable that the BOM tends to perform particularly poorly with less than ten training data points, especially if there are many features to be learned.
The reason for this is that the SSM needs at least two training data points that share all values except the scale-out in order to learn the scale-out to speed-up behavior.
If no such group of training data points is available, this SSM, which is based on a third-degree polynomial regression, can return gravely incorrect results.

\subsection{Discussion}



Concluding our evaluation, one can establish that the accuracy of individual models depends on both the amount of training data and its characteristics, like dimensionality and distribution within the feature space.
Some models can also have decreased effectiveness in large extrapolations, which is typical for tree-based models like the GBM\@.
Which of these models performs best for a given job therefore changes as more runtime data gets generated and new training data becomes available.
Here, our strategy of using cross-validation to choose the most promising model, given the available training data, has shown to be beneficial.

Overall, the C3O predictor exhibits a significant accuracy gain in collaborative scenarios with globally collected context-aware training data compared to local-only training data.
With a sufficient amount of well-distributed global training data, it can keep the mean absolute error below 3\% for each of the jobs in the distributed dataflow runtime dataset.

When only local training data is available, our predictor still regularly performs comparatively well.
We therefore have reason to believe that our novel cluster configurator can also benefit single organizations, perhaps with their own dedicated bare-metal clusters, in finding suitable scale-outs for their jobs.

Naturally, in situations where a user has not yet generated local runtime data, the user is expected to benefit from any globally available data.

\section{Related Work}\label{sec:RELATED_WORK}
Our system aims to be applicable to more than one data processing system, which is why we devised a black-box approach for performance prediction.
This section consequently discusses related black-box approaches to runtime prediction and cluster configuration.

\subsection{Iterative Search-Based}

Some approaches configure the cluster iteratively through profiling runs, attempting to find a better configuration at each iteration, based on runtime information from prior iterations.
They finally settle on a near-optimal solution once it is expected that further searching will not lead to significant enough benefit to justify the incurred overhead~\cite{cherrypick, hsu2018micky, hsu2018arrow, hsu2018scout, tuneful2020}.

For instance, \emph{CherryPick}~\cite{cherrypick} tries to directly predict the optimal cluster configuration, which best meets the given runtime targets.
The search stops once it has found the optimal configuration with reasonable confidence. This process is based on Bayesian optimization.

Another example is \emph{Micky}~\cite{hsu2018micky}, one of three closely related approaches~\cite{hsu2018micky, hsu2018arrow, hsu2018scout}.
It tries to reduce the profiling overhead by doing combined profiling for several workloads simultaneously.
For limiting overhead, it further reformulates the trade-off between spending time looking for a better configuration vs.\ using the currently best-known configuration as a multi-armed bandit problem.

With \emph{Tuneful}~\cite{tuneful2020} an approach exists that combines incremental sensitivity analysis and Bayesian optimization to identify near-optimal configurations. Its setup allows the tuning to be done online and proves to significantly reduce the exploration costs for finding close-to-optimal configurations.

Compared to these approaches, our solution avoids profiling and its associated overhead.

\subsection{Performance Model-Based}

Other approaches use runtime data to predict the scale-out and runtime behavior of jobs.
This data is gained either from dedicated profiling or previous full executions~\cite{bell, ernest, verbitskiy2018cobell, ellis, koch2017smipe, deepRM2016, perforator, scheinert2021bellamy}.

For instance, \emph{Ernest}~\cite{ernest} trains a parametric model for the scale-out behavior of jobs on the results of sample runs on reduced input data.
This works out well for programs exhibiting an intuitive scale-out behavior.
Ernest chooses configurations to try out based on optimal experiment design.

Another example is our own previous work \emph{Bell}~\cite{bell}, which automatically chooses via cross-validation between a parametric model ba\-sed on that of Ernest and a non-pa\-ra\-met\-ric model.
Moreover, it can learn the job's scale-out behavior from historical full executions. Contrary to C3O, Bell is designed for non-collaborative use cases as it assumes data from a single job execution context.

Similarly, \emph{PerfOrator}~\cite{perforator} makes use of non-linear regression on profile runs, calibration queries, and analytical framework models. The system is thus capable of predicting the overall resource skyline of jobs, as well as cost and runtime.

A different approach is conducted with \emph{DeepRM}~\cite{deepRM2016}, where the authors employ deep reinforcement learning in order to manage resources directly from experience. Their solution adapts to different conditions and learns promising strategies. However, the utilization of DeepRM is accompanied by assumptions that can not always be assumed to be fulfilled.

The clear disadvantage of all approaches based on dedicated profiling runs to gain training data is the associated overhead in both time and to some extent the cost.
Our proposed system will not rely on profiling runs.
Historical runtime data for a job is not always available within an organization.

Previously, we explored the possibilities and challenges that a collaborative approach to cluster configuration entails~\cite{will2020towards}.
Towards that idea, this paper presents a system and specific models that can utilize runtime data that was generated in various and significantly different contexts.

\section{Conclusion}\label{sec:CONCLUSION}
The goal of this work was to create a new system that is capable of configuring an efficient public cloud cluster for data analytics workloads while fulfilling the users' runtime requirements.
Towards this goal, we designed and evaluated a collaborative system that allows users to share historical runtime data of distributed dataflow jobs.
The runtime data is shared alongside the code of the job and is used to train black-box runtime prediction models which lie at the core of our cluster configuration system.
The runtime predictor of our envisioned system switches dynamically between a selection of suitable runtime prediction models based on expected accuracy in a given situation.

In realistic scenarios, our runtime predictor has shown to have mean absolute prediction errors below 3\%.
Compared to state of the art systems, our runtime predictions can be significantly more accurate by learning from training data that was generated by other users under consideration of their different execution contexts.

A limitation to the collaborative approach is that users can only benefit from collaboration on commonly occurring jobs.
With highly customized, user-specific jobs, the users will have to rely on runtime data from their own previous executions.
In the future, we therefore plan to work on quick but effective profiling methods to also assist users in that scenario.

\section*{Acknowledgments}

This work has been supported through grants by the German Ministry for Education and Research (BMBF) as BIFOLD (grant 01IS18025A) and the German Research Foundation (DFG) as FONDA (DFG Collaborative Research Center 1404).

\bibliographystyle{IEEEtran}
\balance
\bibliography{./references}

\end{document}